\documentclass[amsmath,amssymb,showpacs,preprint,noshowpacs,showkeys]{revtex4} 
\usepackage{graphicx}
\def\be{\begin{equation}}
\def\ee{\end{equation}}
\def\D{{\mathcal D}}
\def\ep{\epsilon }

\begin{document}
\title{The stationary SQUID}
\author{Jorge Berger}
\affiliation{Department of Physics, Ort Braude College, 21982 Karmiel, Israel} 
\email{jorge.berger@braude.ac.il}
\begin{abstract}

In the customary mode of operation of a SQUID, the electromagnetic field in the SQUID is an oscillatory function of time. In this situation, electromagnetic radiation is emitted, and couples to the sample. This is a  back-action that can alter the state that we intend to measure.
 A circuit that could perform as a stationary SQUID consists of a loop of superconducting material that encloses the magnetic flux, connected to a superconducting and to a normal electrode. This circuit does not contain Josephson junctions, or any other miniature feature. We study the evolution of the order parameter and of the electrochemical potential in this circuit; they converge to a stationary regime and the voltage between the electrodes depends on the enclosed flux. We obtain expressions for the power dissipation and for the heat transported by the electric current; the validity of these expressions does not rely on a particular evolution model for the order parameter. We evaluate the influence of fluctuations. For a SQUID perimeter of the order of 1$\mu$m and temperature $0.9T_c$, we obtain a flux resolution of the order of $10^{-5}\Phi_0/$Hz$^{1/2}$; the resolution is expected to improve as the temperature is lowered.            

\end{abstract}
\maketitle
\section{Introduction}
Superconducting quantum interference devices (SQUIDs) are the most sensitive available fluxmeters. 
SQUIDs are classified into rf SQUIDs, which consist of a loop interrupted by a single Josephson junction and are monitored at radio frequency, and dc SQUIDs, with two Josephson junctions. In principle, a dc SQUID can really be used as a ``dc" fluxmeter by gradually increasing the current through it, until its flux dependent critical current is reached. In a more practical procedure, on which we will focus here, a fixed current, larger than its critical value, is driven through the SQUID, giving rise to a flux dependent average voltage. This voltage is not constant in time, but rather oscillates at a frequency that follows from the Josephson relation.

Many nanoSQUIDs have been built in recent years, with the purpose of measuring the flux generated by nanoscopic samples---eventually the flux due to a single electron. Recent reviews on nanoSQUIDs are available \cite{Fo,We,Carmine,LevS,KoeRev}.
It would not be possible to make justice to the vast literature on the subject; some examples are Refs. \cite{Werns,Caspar,April,Hazra,Russo,NL,Hao1,Gallop,Zel1,Zel2,SQUIPT,SQUIPT1,SQUIPT2}.

As the size of the sample decreases, greater care has to be taken to avoid disturbing the state of the sample as a result of the measurement. It is therefore important to prevent heat dissipation that can reach the sample. A possible strategy is dispersive magnetometry \cite{LevS,P1,disp}, in which the inductance of the SQUID is measured, and it never switches to the resistive state; another possibility is the voltage mode \cite{Zel1,Zel2}, in which dissipation takes place at a shunt. 
But even if no heat reaches the sample, a disturbance channel remains: the electric field in the dc SQUID oscillates, and therefore emits electromagnetic radiation. And since the inductive coupling between the SQUID and the sample must be tight, this radiation will reach the sample. We therefore expect that there will be situations in which a stationary SQUID will be highly desirable.

\section{our proposal}
\begin{figure}
\scalebox{0.85}{\includegraphics{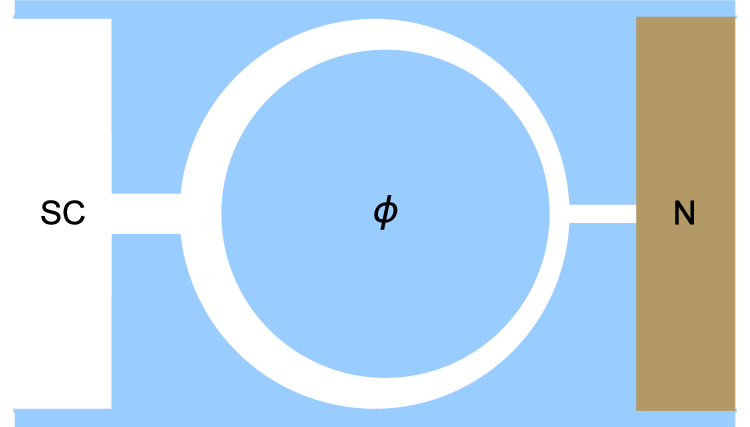}}
\caption{\label{circuit} Circuit designed to measure the flux $\Phi$. The circuit consists of a superconducting loop that encloses the flux, with connections to two electrodes. The electrode at the right is made of a normal metal. 
}
\end{figure}

We have proposed a circuit, as sketched in Fig.~\ref{circuit}, that is predicted to perform as a stationary SQUID \cite{let}. It consists of a loop of superconducting material that encloses the magnetic flux $\Phi $ that we intend to measure, with short wires at the left and the right that connect the loop to ``banks," that serve as electrodes. The connectors and the left bank are made of the same material as the loop, whereas the right bank is a normal metal. A fixed current is driven from the superconducting to the normal bank, and the potential difference between them is measured. This potential difference is a function of $\Phi$, thus providing a measurement of the enclosed flux.

The proposed circuit has close similarity with the normal-metal-–insulator-–superconductor interferometer \cite{Ar}, with the NS-QUID \cite{NS}, and with the superconducting quantum interference proximity transistor \cite{SQUIPT,SQUIPT1,SQUIPT2}.
The new feature is that the connector at the right is in clean contact with the normal metal. The absence of miniature parts in comparison to the entire circuit could result in simpler fabrication at the nanoscale.
Another possibility for a stationary fluxmeter is the ballistic Hall magnetometer \cite{Geim}.

The heuristic argument for SQUID-like performance of this circuit is that, due to fluxoid quantization, current will not only flow from one connector to the other, but a circulating  current will also be present. It follows that the current in one of the branches is expected to exceed half of the total current, increasing the amount of current that has to be transported as normal current. We therefore expect an increase in the voltage between the banks as the enclosed flux deviates from an integer number of quantum fluxes.
In order to gain intuition concerning stationarity, we should first clarify what is the origin of oscillations in a standard SQUID above its critical current: there is a different electric potential at each electrode, leading to different evolutions of the phases of the order parameter at each electrode,  thus creating gradients that periodically force phase slips at some point in the wire. In our case, the order parameter vanishes at the normal bank, so that its phase becomes meaningless, and oscillations will not necessarily occur.

The current between the banks should not be too large, since in this case a considerable fraction of the loop becomes essentially normal, making the circuit insensitive to the flux. On the other hand, the current ought to be sufficiently large to cause the order parameter to vanish at a point in the loop when the flux becomes half a quantum. This feature enables a continuous passage between consecutive winding numbers, thus avoiding hysteresis. 

\section{basic model}
For simplicity, we consider a quasi-1D circuit, and the position dependence of all physical quantities will be completely determined by the arc length. In this section we assume that thermal contact is sufficiently effective to take heating away, so that the entire circuit is kept at a uniform temperature $T$; the expected influence of Joule heating and of supercurrent heat transport will be examined in Sec.~\ref{hot}. Thermal fluctuations will be considered in Sec.~\ref{ThN}.

We choose a gauge with time-independent vector potential, no tangential vector potential along the connectors, and uniform tangential vector potential $A$ along the loop. We write $\ep =1-T/T_c$, where $T_c$ is the critical temperature of the superconductor, and  denote by $\varphi$ the electrochemical potential. The unit of length will be denoted by $x_0$ , by $t_0$ the unit of time, by $\varphi_0$ the unit of voltage, by $A_0$ the unit of vector potential, and by $j_0$ the unit of the current density. We take
\begin{equation}
x_0=\xi(0)\;,\;\;t_0=\frac{\pi\hbar }{8k_BT_c}\;,\;\;A_0=\frac{\hbar c}{2e\xi(0)}\;,\;\;\varphi_0=\frac{4k_BT_c}{\pi e}\;,\;\;
j_0=\frac{4\sigma  k_BT_c}{\pi e\xi (0)}\;,
\label{units}
\end{equation}
where $\xi (0)$ is the coherence length at $T=0$, $k_B$ is Boltzmann's constant, $e$ is the electron charge, and $\sigma $ is the normal conductivity.

A model that can be justified as long as there is local equilibrium is the Kramer--Watts-Tobin model \cite{Kopnin,KWT1,KWT2}, 
which takes into account the different relaxation times of the absolute value and of the phase of the order parameter. In our units, the evolution equations can be written as
\begin{eqnarray}
\!\!\partial _t\psi\!\! &=&h^{-1}[\D^2 +(w'/w) \D ]\psi-i\varphi\psi  \nonumber\\
&+&\!\!
h\psi\{\ep  -  \left| \psi  \right|^2-u\tau_{\rm in}^2{\rm Re}(\psi^*[\D^2 +(w'/w) \D]\psi)\} ,
\label{Eulit}
\end{eqnarray}
and
\begin{equation}
\partial _x\varphi  = u\, {\rm Im}\left( {\psi ^* \D\psi } \right) - j\;.
\label{Ohm}
\end{equation}
Here $\psi$ is the order parameter, with normalization imposed by Eq.~(\ref{Eulit}), $\partial_t$ and $\partial_x$ denote partial differentiation with respect to the time $t$ and to the arc length $x$ along the wire, 
\be
h=(1+u\tau_{\rm in}^2|\psi |^2)^{-1/2} \;,
\label{hh}
\ee
where $u$ is the ratio between the relaxation times of $\psi$ and of the current density $j$, $\tau_{\rm in}$ is the inelastic collision time, 
$\D$ is the operator $\D=\partial_x-i A$, and $w(x)$ is the cross section of the superconducting wire at position $x$. 
The cross section of the left connector will be denoted by $w_{\rm S}$, the cross section of the right connector will be denoted by $w_{\rm N}$, and along the branches of the loop we will take $w(x)$ as a function that changes linearly from $w_{\rm S}$ to $w_{\rm N}$.
$\varphi$ is the potential felt by quasiparticles, that can be measured by means of a normal probe.
Along the branches of the loop, $j$ has to be set as the local current density in the corresponding branch.
Equation (\ref{Eulit}) was worked out in \cite{SCB}, the term for nonuniform cross section was taken from \cite{Giles}, and Eq.~(\ref{Ohm}) is Ohm's law.
The values of $\tau_{\rm in}$ have been tabulated \cite{Kap,scatt} for many materials.
Following \cite{koby}, we denote by $4L$ the perimeter of the loop, and therefore the flux that it encloses is $4LA$. 
Accordingly, we set $A=0$ along the connectors and along the branches of the loop $A=\pm\pi |\Phi|/2L\Phi_0$, where $\Phi_0=\pi\hbar c/e$ is the quantum of flux. 

The boundary conditions require continuity of the order parameter and assume equilibrium at the electrodes. The potential at the superconducting electrode is taken as zero, and the order parameter as $\psi_{\rm S}=\ep^{1/2}$. At the normal electrode, the order parameter is required to vanish. 
At the contacts between the ring and the connectors, we require continuity of the potential and of the order parameter. Charge conservation requires that the sum of total currents along the branches of the loop has to equal the current along the connectors, but assuming that decay or formation of Cooper pairs requires a finite volume, and following \cite{DG,rev}, we impose the stronger constraint $\sum_{n=1}^3\pm\D\psi_n=0$, where $n$ stands for each wire that meets at the contact, we have taken the case that the three wires have the same cross section at the contact, and the sign depends on whether the current is entering or leaving the contact.

Since the order parameter has to be very small near the normal electrode, normal currents and voltage drop will always be present, and stationarity cannot be assumed a priori.

Equations (\ref{Eulit})-(\ref{hh}) are invariant under the gauge transformation $A\rightarrow A+C$, $\psi\rightarrow\exp (iCx)\psi$, where $C$ is constant, but $C$ has to be an integer multiple of $\pi/2L$ in order to obey single valuedness. It follows that any physical property of the circuit is periodic in $\Phi$, with periodicity $\Phi_0$. In addition, switching the sign of $\Phi$ amounts to exchanging the branches of the circuit, so that the potential difference between the electrodes has to be an even function of $\Phi$.

In the limit $\tau_{\rm in}^2|\psi |^2\ll 1$, the Kramer--Watts-Tobin model reduces to the time-dependent Ginzburg--Landau (TDGL) model. Although unrealistic, it is a valuable tool to estimate scaling with length, since in this limit equations (\ref{Eulit})-(\ref{hh}) become invariant under the transformation $x\rightarrow L'x$, $L\rightarrow L'L$, $A\rightarrow L'^{-1}A$, $t\rightarrow L'^2t$, $\psi\rightarrow L'^{-1}\psi$, $\varphi\rightarrow L'^{-2}\varphi $, $\ep \rightarrow L'^{-2}\ep $ and $j\rightarrow L'^{-3}j$, since each of the terms in Eqs.\ (\ref{Eulit}) and (\ref{Ohm}) is multiplied by $L'^{-3}$. The boundary conditions are also invariant under this transformation, provided that the ratios among the lengths of the connectors and the perimeter of the loop are kept unchanged.
Therefore, in this limit it suffices to study a single value of $L$, and the solutions for any other value are obtained by scaling. We note that $L^2\varphi$, $\Phi$ and $L\ep^{1/2}/\xi (0)= L/\xi (T)$ are invariant under this scaling. This scaling breaks down if $L$ is not large in comparison to the average diffusion distance between inelastic collisions \cite{KWT1,KWT2}.

Equations (\ref{Eulit})-(\ref{hh}) were solved numerically, as described in \cite{let}. In the case $w_{\rm S}=w_{\rm N}$ we studied the range
$L^2/\xi^2\alt 10$. The solutions that we found in this range always converged to a stationary regime. We looked for potential differences that are continuous functions of $\Phi$, but the winding number of $\psi$ around the loop cannot change continuously unless $\psi$ vanishes at some point. For this goal, sufficiently large currents are required. Currents that lead to appropriate behavior were found empirically.

\begin{figure}
\scalebox{0.85}{\includegraphics{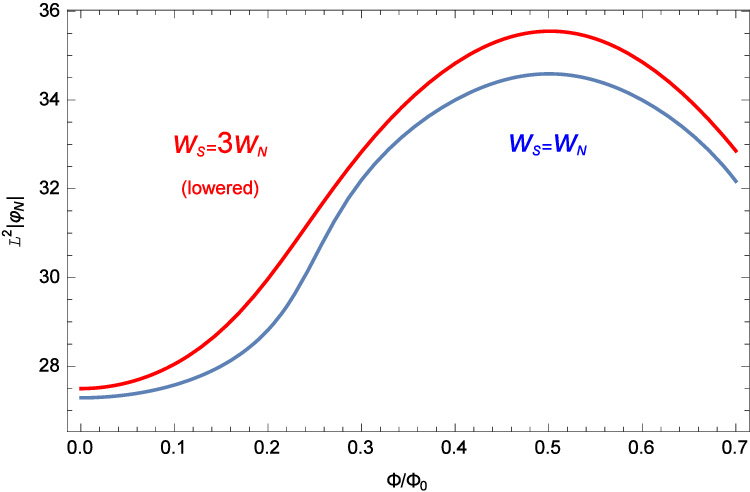}}
\caption{\label{V120} 
Potential difference between the electrodes as a function of the flux for temperature $(1-10\xi^2(0)/L^2)T_c$. $\varphi_{\rm N}$ is the elecrochemical potential at the normal electrode. The blue line is for $w_{\rm S}=w_{\rm N}$ and the red line for $w_{\rm S}=3w_{\rm N}$. In all cases the perimeter of the loop is $4L$, the length of each connector is $0.08L$, and $\tau_{\rm in}=0$. For $w_{\rm S}=w_{\rm N}$ the current density along the connectors is $120j_0\xi^3(0)/L^3$; for $w_{\rm S}=3w_{\rm N}$, the current density along the left connector is $75j_0\xi^3(0)/L^3$. For the purpose of comparison between the two curves, the curve for $w_{\rm S}=3w_{\rm N}$ was lowered by $15
\xi^2(0)\varphi_0$.}
\end{figure}

The lower curve in Fig.~\ref{V120} shows the voltage between the electrodes as a function of $\Phi$ in the TDGL limit, for $w_{\rm S}=w_{\rm N}$, $L^2\ep =10$ and $L^3j=120$. The considered current density is about twice the nominal ``critical current density'' $2u(\ep /3)^{3/2}\approx 70L^{-3}$. Due to periodicity and symmetry of $\varphi_{\rm N}(\Phi )$, the interval $0\le\Phi\le 0.5\Phi_0$ would suffice to describe this function for arbitrary flux, but a larger interval is presented in order to show continuity. For $\Phi =0.5\Phi_0$, the numeric value of $|\psi |$ that we obtained at the right extreme of the loop is of the order of $3\times 10^{-3}\psi_{\rm S}$; this value decreases if a denser computational grid is used. These and the following results were obtained for connectors of length $0.08L$.
 
The upper curve in Fig.\ \ref{V120} shows the voltage between the electrodes as a function of $\Phi$ for $w_{\rm S}=3w_{\rm N}$, $L^2\ep =10$ and $L^3j=75$ (the current density at the connector to the normal electrode is three times larger). In order to enhance visibility, this curve has been lowered by 15 units. Again, no discontinuity is visible in this curve.

\begin{figure}
\scalebox{0.85}{\includegraphics{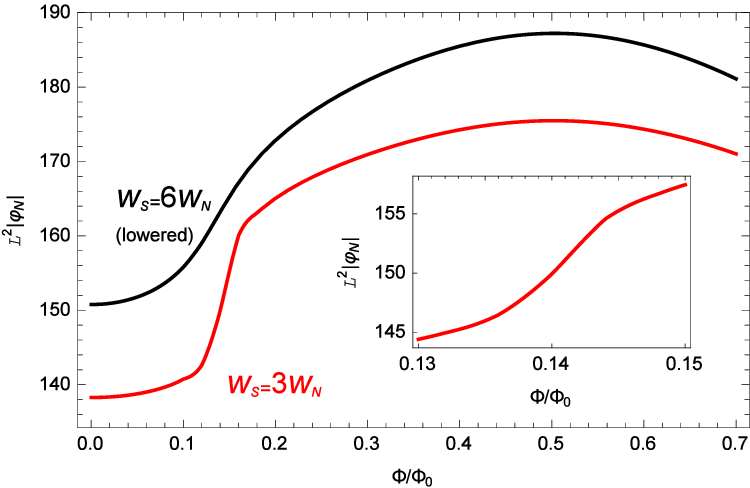}}
\caption{\label{Gam30} 
Potential difference as a function of the flux for temperature $(1-30\xi^2(0)/L^2)T_c$. The red line stands for $w_{\rm S}=3w_{\rm N}$ and $j=320j_0\xi^3(0)/L^3$; the black line, for $w_{\rm S}=6w_{\rm N}$ and $j=230j_0\xi^3(0)/L^3$. For the purpose of comparison between the two cases, the curve for $w_{\rm S}=6w_{\rm N}$ was lowered by $40\xi^2(0)\varphi_0$. The length ratios are the same as in Fig.~\ref{V120}. The inset shows the region near the inflection point for $w_{\rm S}=3w_{\rm N}$ in an expanded scale.}
\end{figure}

In the case $w_{\rm S}>w_{\rm N}$ we have also looked into larger values of $L^2\ep$. Figure \ref{Gam30} shows the voltage between the electrodes for $L^2\ep =30$, for $w_{\rm S}=3w_{\rm N}$ and for $w_{\rm S}=6w_{\rm N}$, still in the TDGL limit. To enhance visibility, the curve for $w_{\rm S}=6w_{\rm N}$ was lowered by 40 units. 
Close to $\Phi\approx 0.14\Phi_0$ the curve for $w_{\rm S}=3w_{\rm N}$ has a steep slope, but is continuous and reversible. By means of an appropriate bias, a steep slope could serve to attain high flux sensitivity.
We note that, unlike the case of a SQUID based on Josephson junctions, the flux sensitivity is proportional to $L^{-2}$, so that the field sensitivity is independent of $L$.
The trend suggested by Figs.\ \ref{V120}-\ref{Gam30} is that the flux-modulation of the voltage increases with $L^2\ep$, and the slope $d\varphi_{\rm N}/d\Phi $ becomes more uniform when $w_{\rm S}/w_{\rm N}$ increases.

\begin{figure}
\scalebox{0.85}{\includegraphics{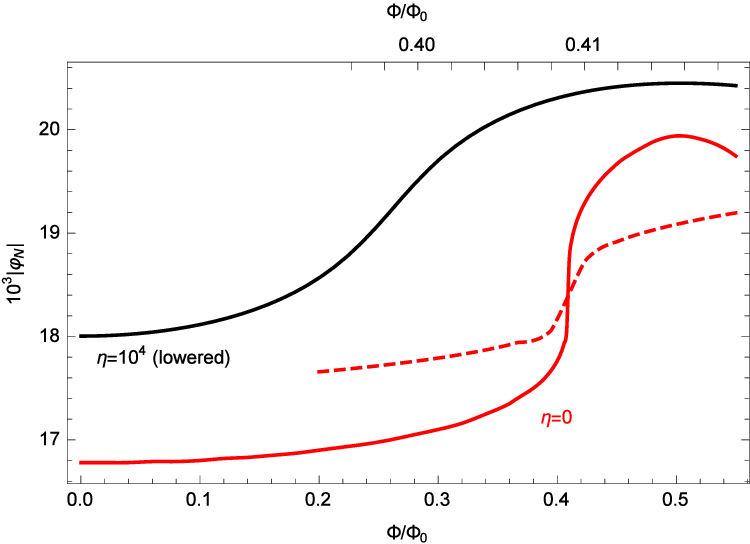}}
\caption{\label{KWf} 
Potential difference as a function of the flux according to the Kramer--Watts-Tobin model, for a circuit with perimeter $400\xi (0)$ and $w_{\rm S}=3w_{\rm N}$, for surrounding temperature $0.997T_c$, current $j=3.2\times 10^{-4}j_0$ and $u\tau_{\rm in}^2=10^4t_0^2$.
The solid red line was evaluated assuming that the current does not produce heating, whereas the black line assumes a local temperature described by Eq.~(\ref{heat}) with $\eta =10^4$. The black line was lowered by 1.5. The dashed line is a horizontal expansion of the solid red line, in the range where it has a steep slope; its vertical scale is the same as for the solid red line, but its horizontal scale is given in the upper axis. }
\end{figure}

Let us now consider the case in which $\tau_{\rm in}^2|\psi |^2$ is not negligible. In this case 
Eqs.\ (\ref{Eulit})--(\ref{hh}) do not obey the scalings that we found for TDGL, forcing us to fix the length of the circuit.
The lower curve in Fig.~\ref{KWf} shows the voltage as a function of the flux for parameters that correspond to the lower curve in Fig.~\ref{Gam30}, but this time the evolution of the order parameter followed Eq.~(\ref{Eulit}) with $u\tau_{\rm in}^2=10^4t_0^2$, which is a typical value for low-$T_c$ superconductors with strong coupling. We note that the main effect that arises from taking $\tau_{\rm in}^2|\psi |^2$ into account is that the range of fluxes for which the voltage rises steeply is shifted towards $\Phi =0.5\Phi_0$. We remark that the flow pattern converges to a stationary situation.

At the inflection point, $d|\varphi_{\rm N}|/d(\Phi /\Phi_0)\approx 0.35$, i.e., the flux sensitivity is $0.35\varphi_0/\Phi_0$. For $T_c\sim 10\,$K, the sensitivity is $\sim 4\times 10^{-4}{\rm V}/\Phi_0$. Assuming that the trend is the same as in Fig.~\ref{Gam30}, higher sensitivities would be found for smaller ratios $w_{\rm S}/w_{\rm N}$ and for smaller lengths.

\section{Heating due to current flow \label{hot}}
We assume that inelastic scattering lengths are short with respect to $L$, so that a local temperature, common to electrons and phonons, can be defined. 
Relaxations of the electric field and of the order parameter lead to local power dissipation. We denote by $W$ the density of power dissipation and by $I_{\rm E}$ the energy flow along the wire.

When a stationary situation is achieved, the power per unit length injected into a short segment of the wire, $-\partial_x I_{\rm E}$, has to equal the power dissipated per unit length, $Ww$, and they both equal  
the power that diffuses away. Heat can either diffuse along the wire or to the substrate, but assuming that the substrate has good thermal conductivity and the thermal healing length $(\kappa d/\alpha)^{1/2}$ \cite{SBT,HP} is also short with respect to $L$, heat will mainly diffuse to the substrate. In the expression of the thermal healing length $d$ is the thickness of the wire, $\alpha$ is the
coefficient of heat transfer to the substrate per unit area, and $\kappa$ is the thermal conductivity of the wire. Relying on this model, we estimate that the local temperature increment is proportional to $W$ and obeys
$\ep =\ep_0-\eta W$,
where $\ep_0$ corresponds to the temperature of the substrate and $\eta =d/\alpha T_c$. 
The density of power dissipation is evaluated in Appendix~\ref{appB}, and we obtain
\be
\ep =\ep_0-\eta \{(\partial_x\varphi )^2+u{\rm Re}[(\partial_t\psi +i\varphi\psi )^*\left( \ep-|\psi |^2+\D^2+(w'/w)\D\right)\psi ]\} \;.
\label{heat}
\ee 

Assuming that heat flow to the substrate equals power dissipation also before the stationary regime is achieved, 
we have solved the system of differential equations (\ref{Eulit})--(\ref{heat}) in order to obtain the functions $\psi (x,t)$, $\varphi (x,t)$, and $\ep (x,t)$. 
The upper curve in Fig.~\ref{KWf} shows the final voltage as a function of the flux when heating is taken into account, assuming Eq.~(\ref{heat}) with $\eta = 10^4$, for the same parameters as for the lower curve.
 We see that the main effect of heating is similar to that of decreasing the ratio $w_{\rm N}/w_{\rm S}$: the steep rise becomes gradual. The similarity between the two cases may be attributed to the reduction of the ability to carry current in the vicinity of the normal electrode. In addition, the size of the voltage modulation decreases, as expected from the fact that there are higher temperatures along the circuit.

As expected, the hottest region in the circuit is found in the connector at the right, close to the branching point. 
Along the branches of the loop, the temperature is almost uniform in most of the circuit and increases gradually in the vicinity of the branching point at the right.
Since the current density is discontinuous at the branching point, the temperature is also discontinuous within the present model.




For values of $\eta$ that are larger (respectively smaller) than $\eta =10^4$, we may expect to obtain a $\varphi_{\rm N}(\Phi )$ curve that is similar to the upper curve in Fig.~\ref{KWf} provided that the surrounding temperature is lowered (respectively raised), so as to yield the same temperature at the right extreme of the loop.

\section{Thermal noise \label{ThN} }
In this section we take thermal noise into account, and investigate to what extent it limits the flux resolution of the proposed device. Thermal noise affects the evolutions of the electrochemical potential and of the order parameter. We will add its influence using a formalism in which length and time are discretized.

In the case of the electrochemical potential, we add the Johnson noise. If $\varphi_{k}$ and $\varphi_{k+1}$ are the electrochemical potentials in two consecutive cells, with a distance $\ell$ between their centers, at periods of time $\tau$ we have to add to $\varphi_{k+1}-\varphi_{k}$ a fluctuating term with gaussian distribution, zero average, and variance
\be
\langle [\Delta (\varphi_{k+1}-\varphi_{k})]^2\rangle =\varphi_0^2 \Gamma_\varphi \frac{T}{T_c}\frac{t_0}{\tau}\;,
\ee
where $\Gamma_\varphi =\pi e^2\ell /\hbar w\sigma$, $w$ stands for the cross section between the two cells, and $\Delta$ stands for the deviation from the value that would be obtained in the absence of fluctuations.

The fluctuating additions to the order parameter have been discussed in previous studies \cite{CM1}. If $\psi_k=|\psi_k|\exp (i\chi_k)$ is the order parameter in cell $k$, then at intervals of time $\tau$ we add to $|\psi_k|$ a fluctuating term with gaussian distribution, average
\be
\langle\Delta |\psi_k|\rangle =\Gamma_\psi\frac{h_k^3}{2|\psi_k|}\frac{T}{T_c}\frac{\tau}{t_0}\;,
\ee
and variance
\be
\langle(\Delta |\psi_k|)^2\rangle -\langle\Delta |\psi_k|\rangle^2=\Gamma_\psi h_k\frac{T}{T_c}\frac{\tau}{t_0}\;,
\ee
where $\Gamma_\psi =\Gamma_\varphi\xi^2(0)/u\ell^2$, $h_k$ is obtained from Eq.~(\ref{hh}) by setting $\psi =\psi_k$, and the cross section $w$ has to be taken as the average in cell $k$. Finally, the addition to the argument is a fluctuating term with gaussian distribution, zero average, and variance
\be
\langle (\Delta \chi_{k})^2\rangle =\frac{\Gamma_\psi}{h_k |\psi_k|^2}\frac{T}{T_c}\frac{\tau}{t_0}\;.
\ee

Obviously, the influence of thermal fluctuations becomes negligible for sufficiently low temperature. Therefore, a more relevant question is whether the signatures encountered in the previous sections are still encountered when the temperature is lowered away from $T_c$. With this in mind, we have studied the influence of thermal noise for $T=0.9T_c$. Accordingly, we reduced the perimeter of the circuit and increased the current density.

\begin{figure}
\scalebox{0.85}{\includegraphics{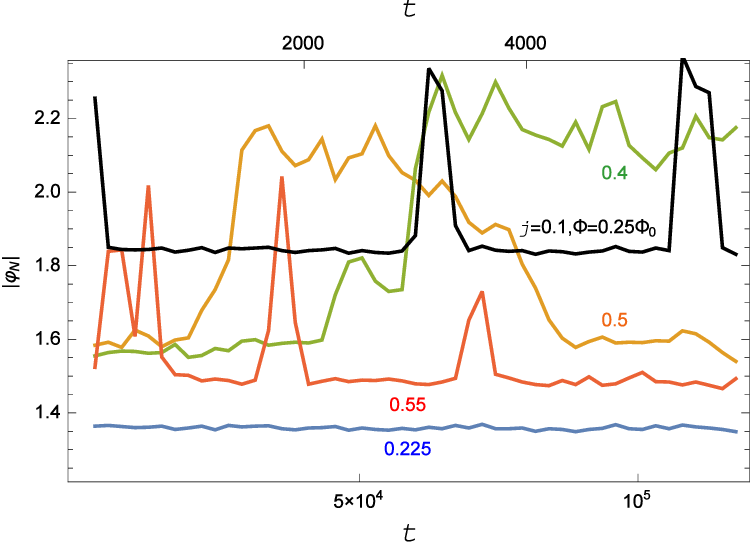}}
\caption{\label{fluctuating} 
Potential difference as a function of time for miscellaneous runs. The colored curves are for $j=0.095j_0$, and the black curve, for $j=0.1j_0$. The number next to each colored curve is the flux in units of $\Phi_0$. For $\Phi =0.225\Phi_0$, for $\Phi =0.55\Phi_0$, and for the black curve, time is shown in the lower horizontal axis and the binning is $2400\,t_0$; for $\Phi =0.4\Phi_0$ and for $\Phi =0.5\Phi_0$, time is shown in the upper axis and the binning is $120\,t_0$. Other parameters: $T=0.9T_c$, $L=20\xi (0)$, $w_{\rm S}=3w_{\rm N}$, $u\tau_{\rm in}^2=10^4t_0^2$, $\eta =0$, $\xi (0)=10\,$nm, $\sigma w_{\rm S}=10^{-6}\Omega^{-1}$cm. For visibility, the curve for $\Phi =0.225\Phi_0$ has been lowered by $0.1\varphi_0$, those for $\Phi =0.4\Phi_0$ and $\Phi =0.5\Phi_0$ have been raised by $0.1\varphi_0$, and the curve for $j=0.1j_0$ has been raised by $0.3\varphi_0$.}
\end{figure}

Figure \ref{fluctuating} shows typical temporal variations of the voltage across the circuit, for some particular runs. 
The colored curves correspond to $j=0.095$.
The curve for $\Phi =0.225\Phi_0$ is typical of the case $0\le\Phi \alt 0.4\Phi_0$; in this case the voltage fluctuates around some average value. On the other hand, for $0.4\Phi_0\alt\Phi \alt 0.6\Phi_0$, as seen in the curve for $\Phi =0.55\Phi_0$, it is clear that fluctuations are not normally distributed. Instead, there are large fluctuations that persist for long periods of time, suggesting that the circuit attains metastable regimes. These large fluctuations can be associated with changes in the winding number of the order parameter around the loop; these changes are also detected in the fraction of the total current that flows through a given branch of the circuit. The curves for $\Phi =0.4\Phi_0$ and for $\Phi =0.5\Phi_0$ have a shorter time span, and depict special situations for which a large fluctuation was present during most of the sampled time.

When the the current density is raised to $j=0.1$, large fluctuations become quite frequent and the flux range where they are present becomes broader. The black curve in Fig.~\ref{fluctuating} shows that these fluctuations are not rare for $\Phi =0.25\Phi_0$.

It might be objected that if the system wanders between two states, it is not strictly in a ``stationary'' regime. Note however that thermal noise is present also in a standard SQUID, in addition to its inherent oscillatory behavior.

\begin{figure}
\scalebox{0.85}{\includegraphics{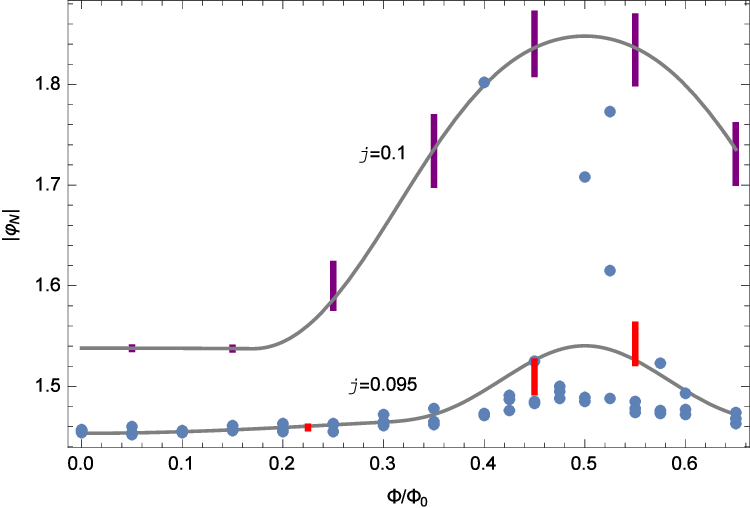}}
\caption{\label{fiofFiFL} 
Average of the potential difference with thermal fluctuations, as a function of the flux. The red lines and the blue dots refer to $j=0.095j_0$; the purple lines refer to $j=0.1j_0$.
The lines were obtained from sampling times that extended during $1.176\times 10^5t_0$, and represent error bars. The dots were obtained from samplings that lasted for $5.88\times 10^3t_0$, and do not show error bars. The gray lines are guides for the eye.
The other parameters are as in Fig.~\ref{fluctuating}.}
\end{figure}

The red lines in Fig.~\ref{fiofFiFL} show average voltages as a function of the flux for $j=0.095$, obtained by monitoring the voltage during a lapse of time in which the winding number changed a handful of times. In order to assess the error, the sampling time was divided into 49 intervals, the average voltage was evaluated for each interval, and then the statistical error was estimated as the standard deviation of the voltage values for these intervals, divided by 7.
As a rough picture of the flux dependence of the voltage, we also evaluated the voltage for several fluxes, for much shorter sampling times. The results are shown as blue dots in Fig.~\ref{fiofFiFL}.

The purple lines in Fig.~\ref{fiofFiFL} show average voltages for $j=0.1$. It appears that in this case our samplings involved a sufficient number of large fluctuations and the results are statistically meaningful. 

We can estimate the flux resolution of our circuit as $0.5\Phi_0\Delta\varphi \sqrt{\Delta t}/|\varphi (0.5\Phi_0)-\varphi (0)|$, where $\Delta\varphi$ is the largest statistical error of $\varphi$ and $\Delta t$ is the sampling time; for the upper curve in Fig.~\ref{fiofFiFL}, the resolution is of the order of $20\Phi_0t_0^{1/2}$. 

\section{Discussion}
We have looked for a superconducting circuit that attains a stationary regime and gives rise to a flux dependent voltage. We have found that this situation is indeed met within a very broad range of parameters. We have not conducted a systematic study to obtain the optimal parameters, but have rather limited ourselves to show a proof of concept for a proposed fluxmeter. Since the working state of this fluxmeter is not oscillatory, it is expected to exert less back action on the measured system.

The circuit considered in Fig.~\ref{fiofFiFL} has a perimeter of $80\xi (0)$, which for $\xi(0)\sim 10\,$nm would be of the order of a micrometer. For flux changes of $0.5\Phi_0$, the voltage changes by about $ 0.4k_BT_c/e$. For $T_c\sim 10\,$K this voltage is of the order of $0.3\,$mV, which should be readily measurable. For $T=0.9T_c$, the flux resolution of this circuit is $\sim 20\Phi_0t_0^{1/2}$, which for $T_c\sim 10\,$K is of the order of $10^{-5}\Phi_0/{\rm Hz}^{1/2}$. 

Maybe the most appealing property of the proposed circuit is the absence of tunnel junctions or any feature that is much smaller than its perimeter. This simplicity should enable the fabrication of particularly small devices.

We have considered temperatures close to $T_c$ just in order to use a theoretically simple model, in which the superconducting state can be totally described by means of the order parameter. 
However, from the practical point of view, we expect that low temperatures will offer a better performance: lower thermal noise and reduced danger of uncontrolled hot spots. In addition, with all other parameters taken equal, 
comparison of Figs.\ \ref{V120} and \ref{Gam30} suggests that lower temperatures would yield larger signals and steeper inflection points. Likewise, we have considered a current bias operation for simplicity, but a voltage bias operation would be advantageous to limit heat dissipation, that is crucial when the sample has to be kept at a very low temperature. We have considered relatively large current biases in order to permit a continuous passage between consecutive winding numbers for fluxes equal to an integer plus half number of quanta; by means of a flux locked operation this feature could become unnecessary, enabling us to use smaller currents and thus obtain larger responsivities. 
\begin{acknowledgments}

Numeric evaluations were performed using computer facilities of the Technion---Israel Institute of Technology. The author has benefited from remarks by Michel Devoret and from correspondence with Hugues Pothier.

\end{acknowledgments}

\appendix

\section{Energy balance in a wire\label{appB}}
Our analysis is an adaptation of Schmid's seminal work \cite{Schmid}. Schmid's analysis was intended to deal with the case of TDGL, but we will keep our expressions in a form that is valid for whatever evolution of the order parameter.

We initially use the gauge in which the scalar potential $\varphi$ is zero. In the units defined in Eq.~(\ref{units}), the Ginzburg--Landau energy of a segment of 1D wire of length $dx$ is
\be
F_{\rm GL}=\frac{u}{2} (-\ep |\psi |^2+\frac{1}{2}|\psi |^4+|\D\psi |^2)wdx\;.
\label{B1}
\ee
The unit of energy is $t_0\varphi_0j_0\xi^2 (0)=2\sigma \hbar k_BT_c\xi (0)/\pi e^2$.

Taking the derivative with respect to time we obtain
\be
dF_{\rm GL}/dt=(u/2) [(\partial_t\psi^*)(-\ep+|\psi |^2+iA\D)\psi +(\partial_{tx}^2\psi^*)\D\psi +i(\partial_tA)\psi^*\D\psi ]wdx+{\rm cc}\;,
\ee
and substituting $w(\partial_{tx}^2\psi^*)\D\psi =\partial_x[w(\partial_t\psi^*)\D\psi ]-(\partial_t\psi^*)[(\partial_xw)\D\psi +w\partial_x(\D\psi )]$, this becomes
\begin{eqnarray}
dF_{\rm GL}/dt=(u/2)&\{& [(\partial_t\psi^*)\left( -\ep+|\psi |^2-\D^2-(w'/w)\D\right)\psi \nonumber \\
&+&i(\partial_tA)\psi^*\D\psi ]w +\partial_x[w(\partial_t\psi^*)\D\psi ]\} dx+{\rm cc}\;.
\end{eqnarray}

We now switch to the gauge in which $A$ does not depend on time. $\partial_tA$ is substituted by $\partial_x\varphi $  and $\partial_t\psi$ by $\partial_t\psi+i\varphi$. We obtain
\begin{eqnarray}
dF_{\rm GL}/dt=&u&\{{\rm Re}[(\partial_t\psi +i\varphi\psi )^*\left( -\ep+|\psi |^2-\D^2-(w'/w)\D\right)\psi ]w \nonumber \\
&-&(\partial_x\varphi ){\rm Im}(\psi^*\D\psi )w
+\partial_x{\rm Re}[w(\partial_t\psi +i\varphi\psi )^*\D\psi ]\} dx\;.
\label{Blast}
\end{eqnarray}

In addition to the Ginzburg--Landau energy, we take into account the electrochemical energy. Assuming electroneutrality, its rate of increment is $wj\partial_x\varphi\, dx$. Adding this quantity to Eq.~(\ref{Blast}) and 
using Eq.~(\ref{Ohm}), the rate of energy change of the segment becomes
\begin{eqnarray}
dF_{\rm total}/dt&=&\{u{\rm Re}[(\partial_t\psi +i\varphi\psi )^*\left( -\ep+|\psi |^2-\D^2-(w'/w)\D\right)\psi ]w \nonumber \\
&-&(\partial_x\varphi )^2 w
+u\partial_x{\rm Re}[w(\partial_t\psi +i\varphi\psi )^*\D\psi ]\} dx\;.
\label{last}
\end{eqnarray}

We finally note that if the volume density of power dissipation is $W$, and the energy flow is $I_{\rm E}$, then this rate of change is $dF_{\rm total}/dt=-(Ww+\partial_xI_{\rm E})dx$. From here and from Eq.~(\ref{last}) we identify
\be
W=(\partial_x\varphi )^2+u{\rm Re}[(\partial_t\psi +i\varphi\psi )^*\left( \ep-|\psi |^2+\D^2+(w'/w)\D\right)\psi ] \,,
\ee
and
\be
I_{\rm E}= -uw{\rm Re}[(\partial_t\psi +i\varphi\psi )^*\D\psi ]\;.
\label{IE}
\ee

The first term in $W$ is the Joule dissipation, and the second term is due to the relaxation of the order parameter. In the limiting case of TDGL, $\partial_t \psi =[\ep-|\psi |^2- i\varphi +\D^2+(w'/w)\D]\psi$ and we recover the result in \cite{Schmid}.

Since $\psi$ and $\varphi$ are continuous, it follows from Eq.~(\ref{IE}) that
the constraint $\sum_{n=1}^3\pm\D\psi_n=0$ not only ensures charge conservation at the contacts between the ring and the connectors, but also energy conservation. 



\begin{thebibliography}{99}
\bibitem{Fo} C.P. Foley and H. Hilgenkamp, Supercond. Sci. Technol. {\bf 22}, 064001 (2009).
\bibitem{We} W. Wernsdorfer, Supercond. Sci. Technol. {\bf 22}, 064013 (2009).
\bibitem{Carmine} C. Granata and A. Vettoliere, Phys. Rep. {\bf 614}, 1 (2016).
\bibitem{LevS} E. M. Levenson-Falk, N. Antler and I. Siddiqi, Supercond. Sci. Technol. {\bf 29}, 113003 (2016).
\bibitem{KoeRev} M. J. Mart\'{i}nez-P\'{e}rez and D. Koelle, {\it NanoSQUIDs: Basics \& recent advances}, Physical Sciences Reviews (De Gruyter, Berlin, in press); arXiv:1609.06182.

\bibitem{Werns} W. Wernsdorfer, D Mailly, A. Benoit, J. Appl. Phys. {\bf 87}, 5094 (2000).
\bibitem{Caspar} C. H. van der Wal, A. C. J. ter Haar, F. K. Wilhelm, R. N. Schouten, C. J. P. M. Harmans, T. P. Orlando, S. Lloyd,  J. E. Mooij, Science {\bf 290}, 773 (2000).
\bibitem{April} J.-P. Cleuziou, W. Wernsdorfer, V. Bouchiat, T. Ondar\c{c}uhu, and M. Monthioux, Nat. Nanotech. {\bf 1}, 53 (2006).
\bibitem{Russo} R. Russo, C. Granata, E. Esposito, D. Peddis, C. Cannas, and  A. Vettoliere, Appl. Phys. Lett. {\bf 101}, 122601 (2012).
\bibitem{Hazra} D. Hazra, J. R. Kirtley and K. Hasselbach, Appl. Phys. Lett. {\bf 104}, 152603 (2014).

\bibitem{NL} A. G. P. Troeman, H. Derking, B. Borger, J. Pleikies, D. Veldhuis, and H. Hilgenkamp, Nano Lett. {\bf 7}, 2152 (2007).
\bibitem{Hao1} L. Hao, J. C. Macfarlane, J. C. Gallop, D. Cox, J. Beyer, D. Drung, and T. Schurig, Appl. Phys. Lett. {\bf 92}, 192507 (2008).
\bibitem{Gallop} L. Hao, J. C. Gallop, D. C. Cox, and J. Chen, IEEE J. Selected Topics in Quantum Electronics, {\bf 21}, 1 (2015).
\bibitem{Zel1} A. Finkler, Y. Segev, Y. Myasoedov, M. L. Rappaport, L. Ne'eman, D. Vasyukov, E. Zeldov, M. E. Huber, J. Martin, and A. Yacoby, Nano Lett., {\bf 10}, 1046 (2010).
\bibitem{Zel2} D. Vasyukov, Y. Anahory,	L. Embon, D. Halbertal,	J. Cuppens,	L. Neeman, A. Finkler, Y. Segev, Y. Myasoedov, M. L. Rappaport,	M. E. Huber, and Eli Zeldov, Nat. Nanotech. {\bf 8}, 639–644 (2013).

\bibitem{SQUIPT} F. Giazotto, J. T. Peltonen, M. Meschke, and J. P. Pekola, Nature Phys. {\bf 6}, 254 (2010).
\bibitem{SQUIPT1} R. N. Jabdaraghi, M. Meschke, and J. P. Pekola, Appl. Phys. Lett. {\bf 104}, 082601 (2014).
\bibitem{SQUIPT2} A. Ronzani, C. Altimiras, and F. Giazotto, Phys. Rev. Applied {\bf 2}, 024005 (2014).

\bibitem{P1} A. Lupa\c{s}cu, E. F. C. Driessen, L. Roschier, C. J. P. M. Harmans, and J. E. Mooij, Phys. Rev. Lett. {\bf 96}, 127003 (2006).
\bibitem{disp} M. Hatridge, R. Vijay, D. H. Slichter, J. Clarke, and I. Siddiqi, Phys. Rev. B {\bf 83}, 134501 (2011).

\bibitem{let} J. Berger, J. Phys.: Condens. Matter {\bf 29}, 29LT01 (2017).
\bibitem{Ar} K. Yu. Arutyunov and T. T. Hongisto, Phys. Rev. B {\bf 70}, 064514 (2004).
\bibitem{NS} S. Gu\'{e}ron, doctoral thesis submitted to University of Paris 6 (1997), http://iramis.cea.fr/drecam/spec/Pres/Quantro/static/publications/phd-theses/index.html;
H. Pothier, S. Gu\'{e}ron, D. Esteve, and M. H. Devoret, Phys. Rev. Lett. {\bf 73}, 2488 (1994). See also V. T. Petrashov, V. N. Antonov, P. Delsing and T. Claeson, Phys. Rev. Lett. {\bf 74}, 5268 (1995).
\bibitem{Geim} A. K. Geim, S. V. Dubonos, J. G. S. Lok, I. V. Grigorieva, J. C. Maan, L. T. Hansen and P. E. Lindelof, Appl. Phys. Lett. {\bf 71}, 2379 (1997).


\bibitem{Kopnin}N.B. Kopnin, {\it Theory of Nonequilibrium Superconductivity} (Clarendon Press, Oxford, 2001).
\bibitem{KWT1} L. Kramer and R. J. Watts-Tobin, Phys. Rev. Lett. {\bf 40}, 1041 (1978). 
\bibitem{KWT2} R. J. Watts-Tobin, Y. Kr\"{a}henb\"{u}hl and L. Kramer, J. Low Temp. Phys. {\bf 42}, 459 (1981).
\bibitem{SCB} J. Berger, Phys. Rev. B {\bf 92}, 064513 (2015).
\bibitem{Giles} G. Richardson and J. Rubinstein, Proc. R. Soc. Lond. A {\bf 455}, 2549 (1999).

\bibitem{Kap} S. B. Kaplan, C. C. Chi, D. N. Langenberg, J. J. Chang, S. Jafarey, and D. J. Scalapino, Phys. Rev. B {\bf 14}, 4854; erratum Phys. Rev. B {\bf 15}, 3567 (1977). 
\bibitem{scatt} L. Parlato, R. Latempa, G. Peluso, G. P. Pepe, R. Cristiano, and R. Sobolewski, Supercond. Sci. Technol. {\bf 18}, 1244 (2005).
\bibitem{koby} J. Rubinstein, P. Sternberg, and Q. Ma, Phys. Rev. Lett. {\bf 99}, 167003 (2007).

\bibitem{DG} P. G. de Gennes, C.\ R.\ Acad.\ Sci.\ Paris II {\bf 292}, 279 (1981).
\bibitem{rev} M. Kato and O. Sato, Supercond. Sci. Technol. {\bf 26}, 033001 (2013).

\bibitem{SBT} W. J. Skocpol, M. R. Beasley, and M. Tinkham, J. Appl. Phys. {\bf 45}, 4054 (1974).
\bibitem{HP} D. Hazra, L. M. A. Pascal, H. Courtois, and A. K. Gupta, Phys. Rev. B {\bf 82}, 184530 (2010).



\bibitem{CM1} J. Berger, J. Phys.: Condens. Matter {\bf 23}, 225701 (2011).

\bibitem{Schmid} A. Schmid, Phys. Kondens. Materie {\bf 5}, 302 (1966). Note that in Eq.~(12.3) in Ref.~\onlinecite{Kopnin}, $\hbar$ was set as 1.
\end{thebibliography}
\end{document}